\documentclass[prl,twocolumn,superscriptaddress,showpacs,amsmath,amssymb]{revtex4}

\usepackage{bm}

\usepackage{graphicx}
\usepackage{latexsym}
\usepackage{amsmath}
\usepackage{amssymb}
\usepackage{amsfonts}
\usepackage{color}
\usepackage{bm}
\usepackage{verbatim}

\begin{document}

\title{Unusual mechanism of vortex viscosity generated by mixed normal modes in superconductors with broken time
reversal symmetry}

 \author{Mihail Silaev}
 \affiliation{O. V. Lounasmaa Laboratory, P.O. Box 15100, FI-00076
 Aalto University, Finland}
 \affiliation{Institute for Physics of
 Microstructures RAS, 603950 Nizhny Novgorod, Russia.}
 \affiliation{Department of Theoretical Physics, The Royal
 Institute of Technology, Stockholm, SE-10691 Sweden} \affiliation{
 Department  of Physics, University of Massachusetts Amherst, MA
 01003 USA }
 \author{Egor Babaev}
 \affiliation{Department of Theoretical Physics, The Royal
 Institute of Technology, Stockholm, SE-10691 Sweden} \affiliation{
 Department  of Physics, University of Massachusetts Amherst, MA
 01003 USA }

\date{\today}

\begin{abstract}
 We show that under certain conditions multiband
superconductors with broken time-reversal symmetry have a new
vortex viscosity-generating mechanism which is different from that
in conventional superconductors. It appears due to the existence
of mixed superfluid phase-density mode inside vortex core. This
new contribution is dominant near the time reversal symmetry
breaking phase transition. The results could be relevant for three
band superconductor $Ba_{1-x}K_{x}Fe_2As_2$.
\end{abstract}

 \maketitle

Recent discoveries of many novel multiband superconducting
compounds have motivated the current quest of  theoretical
understanding of their basic  properties. Especially strong impact
has the recent discovery of iron based superconductors
\cite{FeAs}. Namely it was discussed in \cite{Ng,StanevTesanovic}
that such superconductors can break time reversal symmetry because
these system can have frustrated ground state values of the order
parameter phase differences in different bands $\theta_{ik} =
\theta_i-\theta_k \neq\pi n$. In that case a ground state has a
broken
 time reversal symmetry (BTRS) which is associated with the complex conjugate of the order parameter
 $\psi\rightarrow \psi^*$.
Therefore such superconductors break $U(1)\times Z_2$ symmetry
\cite{Johan}. Physically this implies existence of persistent
"Josephson current" between the three bands which is different for
two ground states. It was recently demonstrated  that such physics
very likely occurs in
 strongly hole doped $Ba_{1-x}K_{x}Fe_2As_2$ \cite{Chubukov}.
Alternatively the other scenarios of time reversal symmetry
breakdown in iron-based superconductors have been discussed
recently \cite{Other}.

In this kind of BTRS state there appear new phenomena which are
absent in conventional and even extended $s$ wave multiband
superconductors. Indeed it has been shown to support a new kind of
topological defects - $CP^2$ skyrmions \cite{Babaev},
 Leggett's mode which becomes massless at the $Z_2$ phase transition \cite{Lin},
as well as mixed phase-density collective modes in $U(1)\times
Z_2$ state \cite{Johan}. Moreover even in the $U(1)$ frustrated
systems, time reversal symmetry breakdown can occur inside vortex
excitations\cite{Johan}. Beyond the mean field approximation and
for sufficiently strong frustration of interband interactions such
systems can have an unusual normal state
 which breaks $Z_2$ symmetry as a precursor to
 a superconducting phase
transition\cite{BabaevSudbo}.

Since the system can break time reversal symmetry at certain
doping \cite{Chubukov}, the superconducting state in the immediate
vicinity of the time reversal symmetry breaking phase transition
should be very interesting because of the existence of a diverging
length scale associated with the $Z_2$ symmetry breakdown. In this
paper we show that the BTRS superconducting state with vortices
has highly unusual thermodynamic and transport properties near the
$Z_2$ symmetry breaking transition. The peculiarities of vortex
state can be helpful to obtain experimental identification of BTRS
superconductivity in particular compounds.

We employ the three band Ginzburg-Landau (GL) expansion of the
free energy density
\begin{eqnarray}\label{Eq:EnergyDensity1} \nonumber
 f=\sum_{k=1}^3 \left[ \alpha_k|\psi_k|^2+\frac{\beta_k}{2}|\psi_k|^4
    +g_k\left\vert\left(\nabla+i\bf{A}\right)\psi_k\right\vert^2
    \right]+\\
    \sum_{i,k=1}^3 \gamma_{ik} \psi_i\psi_k^*+c.c. + |\nabla\times\bf A|^2/8\pi.
\end{eqnarray}
Here, $\psi_k$ are the order parameters  in each band labelled by
band index $k=1,2,3$ and the second term is interband Josephson
coupling energy characterized by interband coupling constants
$\gamma_{ik}$. The field ${\bf A}$ is vector potential. For formal
microscopic justification of multiband GL functionals see
\cite{silaev}, GL expansion for three band BTRS superconductor was
recently studied in detail in\cite{Chubukov} where it was shown
that the doping level $x$ in $Ba_{1-x}K_xFe_2As_2$ determines the
interband pairing interaction between electron and hole pockets
$u_{he}$. The relation to our parameters is following
$\gamma_{13}=u_{he}$ and $\gamma_{12}= \gamma_{23}=u_{hh}$ which
is the interaction between hole pockets. Such an expansion may
contain also other terms which however will not change
quantitatively conclusions of this paper, thus we choose to work
with the minimal model.

 First we investigate the equilibrium vortex structures in three
 band superconductor.
 We substitute the order parameters to GL equation in
the form $\psi_k=\Delta_k e^{i\theta_k}$ where $\Delta_k$ is real
and separate the real and imaginary parts introducing the gauge
invariant superfluid velocities
${\bf{Q_k}}={\bf{A}}+\nabla\theta_k$.
 It should be noted that even when
a ground state has only $U(1)$ broken symmetry, the GL model
(\ref{Eq:EnergyDensity1}) allows for topological excitations with
phase differences of order parameter components
 $\theta_{ik} = \theta_i-\theta_k \neq\pi n$  \cite{Johan}.
 In the particular case of axially symmetric single vortex in
 BTRS phase this results in the radial dependence of the order parameter phases
 $\theta_{ik}=\theta_{ik} (r)$
 (we assume that vortex center is at the origin $r=0$).

 Thus the additional degree of freedom due to the frustrated
phase difference in three component system allows for a
 static {\it mixed phase-density mode} which appears inside
 vortex cores in BTRS phase. To explore its impact on the vortex physics
 we employ the minimal model which in particular describes possible BTRS
 transition to the $s+is$ state in $Ba_{1-x}K_xFe_2As_2$
 \cite{Chubukov}.

Consider a single vortex in three component
 superconductor described by GL functional (\ref{Eq:EnergyDensity1})
 with  $\alpha_1=\alpha_3=\alpha$ and $\beta_1=\beta_3=\beta$,
 $g_1=g_3=g$. To study the modification of vortex properties
 during  he BTRS transition we fix the values
of $\gamma_{12}=\gamma_{23}=\gamma$ and vary $\gamma_{13}$ which
models the electron-hole interaction determined by the
 level of doping $x$ in $Ba_{1-x}K_xFe_2As_2$ compound\cite{Chubukov}.
  Qualitatively our conclusions will however be valid also for a non-symmetric set of
  couplings.

 For such choice of GL coefficients we will use an
  ansatz for vortex solutions $\theta_{12}=\theta_{23}=\theta (r)$ and
 $\Delta_1=\Delta_3=\Delta(r)$. The GL equations in this case read
  \begin{eqnarray}
  \nonumber
&g\left[\nabla^2_r \theta + 2 (\ln \Delta)^\prime_r
\theta^\prime_r\right]
   + \frac{\gamma\Delta_2}{\Delta}\sin\theta+
   \gamma_{13}\sin(2\theta)=0&\\
  \label{Eq:System}
&\left[g_2\left(\nabla^2_r-\bf{Q_2^2}\right)-\alpha_2-\beta_2\Delta_2^2\right]\Delta_2=
    2\gamma \Delta \cos\theta & \\ \nonumber
 &\left[g\left(\nabla^2_r-\bf{Q^2}\right)-\alpha- \gamma_{13}  \cos(2\theta) -\beta\Delta^2\right]\Delta=\gamma\Delta_2
 \cos\theta &
\end{eqnarray}
where ${\bf{Q}^2}=(A+1/r)^2+ \theta^{\prime 2}_r$ and
${\bf{Q}^2_2}=(A+1/r)^2$. For the vector potential we use a radial
gauge ${\bf A}=A(r)(-\sin\alpha, \cos\alpha)$.

\begin{figure}[!h]
 \centerline{\includegraphics[width=1.0\linewidth]{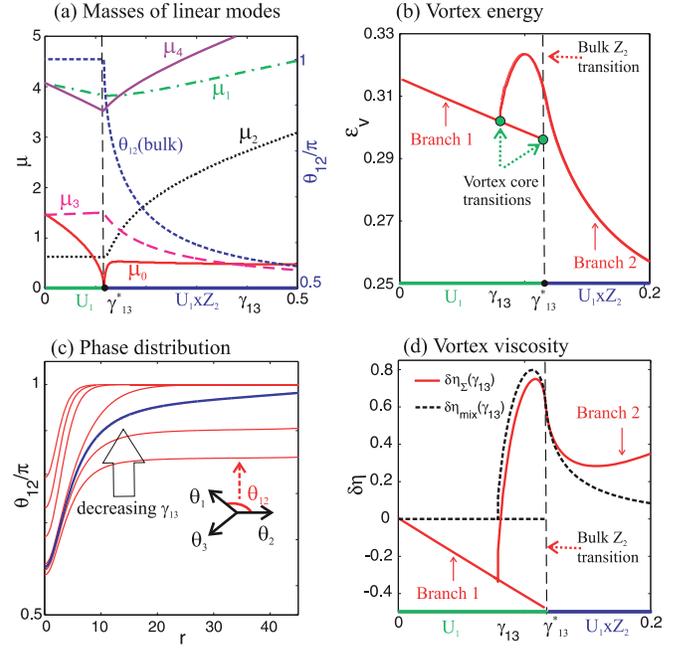}}
 \caption{\label{Fig:order1}
  (a) Masses of the asymptotic mixed modes of the system
 (\ref{Eq:System}). The GL parameters are
 $\alpha_1=\alpha_2=\alpha_3=-0.5$, $\beta_1=\beta_2=\beta_3=10$,
 $\gamma_{12}=\gamma_{23}=0.2$, $g_2=5$ and $g=0.1$.
  The modes $\mu_{0,1,2}$ corresponds to
 symmetric excitations with $\Delta_1=\Delta_3$. The modes $\mu_{3,4}$
 break this symmetry. By dashed
 blue line in (a) the ground state (bulk) phase difference is shown.
 (b) Two branches of vortex energy.  Branch 1 corresponds to the
 vortex solutions which do not break time reversal symmetry.
 Branch 2 corresponds to the  solutions with  non-homogeneous relative phase
(i.e. BTRS solutions).  (c) Relative phase distribution
 $\theta_{12}=\theta_{23}=\theta (r)$ inside vortex core corresponding to the Branch 2.
 For decreasing $\gamma_{13}$ one first meets the second order phase transition in bulk where the characteristic  scale
 of variation
 of $\theta_{12}(r)$ is the largest (blue curve). At the critical value of $\gamma_{13}=\gamma^c_{13}$ the
 Branches 1 and 2 merge when the amplitude of $\theta_{12}=\theta_{12}(r)$ decreases to zero as $\sqrt{\gamma_{13}-\gamma^c_{13}}$.
 (d) Vortex viscosity variation $\delta\eta(\gamma_{13})=\eta(\gamma_{13})-\eta(0)$
 (red solid line is total viscosity and black dashed line
 is mixed mode contribution). }
 \end{figure}

 First let us consider the modification of asymptotical
properties of the system (\ref{Eq:System}) far from the vortex
center during the BTRS transition. At small couplings
$\gamma_{13}<\gamma_{13}^*$ the system is in the plain $U(1)$
symmetry breaking state with the relative phase between
superconducting components $\theta_{12}=\pi$.  The critical value
of coupling separating the $U(1)$ and $U(1)\times Z_2$ bulk phases
is given by $\gamma^*_{13}=\gamma\Delta_{02}/2\Delta_0$, where
$\Delta_{0}$ and $\Delta_{20}$ are bulk values of the amplitudes
$\Delta$ and $\Delta_2$. Beyond the threshold
$\gamma_{13}>\gamma_{13}^*$ the time reversal symmetry is broken
so that $0<\theta_{12}<\pi$. This behavior of bulk $\theta_{12}$
is shown in Fig.(\ref{Fig:order1})a by blue dashed curve.

The masses of symmetric mixed modes obtained by the linearization
of the system (\ref{Eq:System}) are shown in
Fig.(\ref{Fig:order1})a as functions of the coupling
$\mu=\mu_{0,1,2}(\gamma_{13})$. In general the three band GL model
(\ref{Eq:EnergyDensity1}) has five distinct mixed modes which are
the fundamental solutions of linearized equations \cite{Johan}.
The masses $\mu_{3,4}$ correspond to non-symmetric modes with
$\Delta_1 (r)\neq \Delta_3 (r)$ and $\theta_{12} (r)\neq
\theta_{13} (r)$.

In $U(1)$ phase when $\gamma_{13}<\gamma^*_{13}$ the mode shown by
red solid line is a pure phase one which is decoupled from order
parameter densities. However in general it can still be excited
inside vortex core due to nonlinearities.  In this case $Z_2$
symmetry can be broken locally in the core but not in the bulk
\cite{Johan}. At the critical point of $Z_2$ symmetry breakdown in
the bulk of the system the mixed mode has zero mass
$\mu_0(\gamma_{13}^*)=0$ \cite{remm2}.

At $Z_2$ critical point, existence of massless mode results in a
power-law localization of vortex-core solutions. Note that in this
case the {\it anharmonism} in Eq.(\ref{Eq:System}) is important.
Here at large $r$ the field deviations from bulk values can be
found in the form of power law expansions (the details of
asymptotic analysis are given in Supplementary material)
$(\tilde{\Delta},\tilde{\Delta}_2,\tilde{\theta})=
(C_\Delta/r^2,C_{\Delta 2}/r^2,C_\theta/r)$.

 Now let us search numerically for vortex solutions of Eqs.
(\ref{Eq:System}) to show that the vortex energy and viscosity
have anomalies at $Z_2$ phase transition. As will be discussed
below, it is important to take into account the following three
circumstances  for the accurate description of this anomalous
behavior at the BTRS phase transition (i) phase-density modes
mixing, (ii) appearance of massless mode and (iii){\it
anharmonism} in Eqs.(\ref{Eq:System}).

 To find possible vortex structures we implement a numerical solution of the full GL system
 (\ref{Eq:System}) supplemented by the Ampere's law for magnetic field.
 To define the boundary conditions for the fields we consider a vortex lattice and thus use the
 circular cell approximation (see also remark \cite{remark}). At
 the boundary of the Weigner-Seits cell $r=r_s$ the fields satisfy
 $\Delta^\prime=\Delta_2^\prime=\theta^\prime=0$ and   $rA+1=0$.
 The former determines the order parameter to be periodic function
 and the latter one provides magnetic
 flux quantization. Also from the first
 of Eqs.(\ref{Eq:System}), it follows that the boundary condition for
 the phase is $\theta^\prime_r (r=0)=0$.

We investigated the vortex structure as function of interband
coupling $\gamma_{13}$ which is determined by the doping level in
$Ba_{1-x}K_xFe_2As_2$ compound\cite{Chubukov}.
 We have found that the relative phases of the order parameter
components in BTRS superconductor $\gamma_{13}>\gamma_{13}^*$
always have non-trivial variation
$\theta_{12}=\theta_{23}=\theta(r)$ in contrast to the usual
time-reversal invariant case.  Examples of phase distributions are
shown in Fig.(\ref{Fig:order1})c for a set of $\gamma_{13}$ values
decreasing towards the vortex core transition point
$\gamma^c_{13}$ which will be discussed below. We will see that
such phase variation produces an additional friction force on the
moving vortices.

For $\gamma_{13}>\gamma_{13}^*$ the vortex solution is unique. Its
energy is shown in Fig.(\ref{Fig:order1})b by solid red curve
denoted as Branch 2. Even at the point of bulk $Z_2$ transition
  the energy remains finite due to the discussed above {\it anharmonism} of the massless
 mixed mode in Eqs.(\ref{Eq:System}). However its contribution provides a peak of vortex energy close to
 $\gamma_{13}=\gamma_{13}^*$ (where this mode becomes massless).

On the other hand time reversal invariant state at
$\gamma_{13}<\gamma_{13}^*$ can support two different vortex
structures. To demonstrate it we note at first that the
Eqs.(\ref{Eq:System}) always have solution with the relative phase
$\theta=\pi$. We find that this solution is stable in $U(1)$
domain. At the same range of parameters the $Z_2$ symmetry can be
broken in the vortex core leading to the non-trivial variation of
$\theta(r)$ with asymptotic boundary condition
$\theta(r\rightarrow \infty) = \pi$. We find numerically that
these vortex structures can coexist at a certain region
$\gamma_{13}<\gamma^*_{13}$ (note that one of the solutions can be
unstable  \cite{remm} in the coexistence region, but this does not
affect  conclusions of this paper). The corresponding branches of
vortex energy are shown in Fig.(\ref{Fig:order1}b). There is a
critical value of interband coupling $\gamma^c_{13}<\gamma^*_{13}$
where the two branches merge. This critical coupling is determined
as the eigenvalue of linearized first equation in the system
(\ref{Eq:System}) which we write in the form of Sturm-Liouville
equation
$   \hat L \tilde{\theta}=
   2\gamma^c_{13} \Delta^2 \tilde{\theta}$
 where $\tilde{\theta}=\pi-\theta $ and
$\hat L= -g\left[\Delta^2\nabla^2_r  + (\Delta^2)^\prime_r
\partial_r\right]
   + \gamma\Delta_2\Delta $
is a hermitian operator.
 This means (see Supplementary Material) that at $\gamma_{13}>\gamma^c_{13}$ the amplitude of relative phase
variation is given by $\tilde{\theta}\sim
\sqrt{\gamma_{13}-\gamma^c_{13}}$ so that the energy difference
between Branch 1 and Branch 2 is linear
$\varepsilon_2-\varepsilon_1\sim \gamma_{13}-\gamma^c_{13}$.

 The obtained BTRS modification of vortex core structure is manifested transport properties determined by
 vortex viscosity.
  To describe a non-equilibrium process of vortex motion,
  we use
  time-dependent GL model (TDGL) (for a review of TDGL approach see e.g.\cite{KopninBook, Dorsey}) generalized to a multiband case
      \begin{equation}        
    \Gamma_k (\partial_t-i \varphi)\psi_k=
    - \frac{\delta}{\delta\psi^*_k} \int f d^3 {\bf r}
    \label{fder}
    \end{equation}
  where $k=1,2,3$,  $\Gamma_k$ are damping constants and $\varphi$ is the potential of a quasistationary electric
  field.
 Choosing Couloumb gauge for the vector potential ${\rm
div}{\bf A}=0$ we obtain the Poisson equation (see Supplementary
material for detailed discussion)
 \begin{equation}\label{Eq:Phi}
 \sigma_n\triangle\varphi = 2 \sum_{k=1}^3\Gamma_k\Delta_k^2(\varphi-\dot{\theta}_k)
 \end{equation}
 where $\sigma_n$ is a normal state electric conductivity. Equation (\ref{Eq:Phi})
  will be employed to calculate the distribution of electric field  generated by a moving vortex.

 Vortex motion introduces
 a distortion of the order parameter and vector potential fields.
 For a slow vortex motion with a given velocity ${\bf U}$ we calculate the time dependence by making
 Galilean transformation ${\bf r}\rightarrow {\bf r}- {\bf U}t$ of equilibrium
 fields so that $\partial_t=- ({\bf U}\cdot \nabla)$.
We now assume that ${\bf U}=U {\bf x}$
 and search for the electrostatic potential in the form
 $\varphi=U[\varphi_\alpha(r) \sin\alpha- \varphi_r(r) \cos\alpha ]$.
 The  resulting equations read
 \begin{eqnarray}\label{Eq:phi-alpha}
 \sigma_n\left(\nabla^2_r-1/r^2\right)\varphi_\alpha=2\sum_{k=1}^3\Gamma_k\Delta_k^2
 \left(\varphi_\alpha-1/r\right) \\
 \sigma_n\left(\nabla^2_r-1/r^2\right)\varphi_r=2\sum_{k=1}^3\Gamma_k\Delta_k^2
 \left(\varphi_r+\theta_k^\prime\right) \label{Eq:phi-r}
 \end{eqnarray}
 where $\theta_k^\prime=\partial\theta_k/\partial r$.
  Note that in Eq.(\ref{Eq:phi-r}) the derivatives $\theta_k^\prime$
 can be expressed through the two functions $\theta_{12}(r)$ and $\theta_{13}(r)$
 using the condition for the radial current to be zero $\sum_k
 g_k\Delta_k^2\theta_k^\prime=0$.
In the circular cell approximation the boundary condition require
the tangential component of the electric field to be zero
 $({\bf E}\cdot {\bf e_\alpha})|_{r=r_s} =0$. Recalling that
 ${\bf \dot{A}}=-U \cos\alpha A^\prime {\bf e_\alpha}$ we obtain at
 $ \varphi_\alpha(r_s)-r_s A^\prime(r_s)=0$ and $\varphi_r(r_s)=0$.
Also from the Eqs.(\ref{Eq:phi-alpha},\ref{Eq:phi-r}) follows that
$\varphi_\alpha (r=0)=\varphi_r (r=0)=0$.

At first we note that Eq.(\ref{Eq:phi-alpha}) coincides with that
for the vortices in time reversal invariant superconductors (see
e.g. \cite{KopninBook,Dorsey}). It determines Bardeen-Stephen
vortex viscosity \cite{BardeenStephen}. The second
Eq.(\ref{Eq:phi-r}) determines qualitatively new part of the
scalar potential which appears due to the phase-density mixed mode
in BTRS superconductor.  The source in the r.h.s. of this equation
is determined by the {\it radial} dependencies of the relative
phase $\theta_{12} (r)$ [example is shown in the
Fig.(\ref{Fig:order1})c].

Consider now electric field distribution generated by a moving
vortex. The electric field can be written as a superposition of
two terms ${\bf E}= {\bf E_\alpha}+{\bf E_r}$ where
 ${\bf E_\alpha}=  U\left[\nabla (\varphi_\alpha \sin\alpha)-\cos\alpha A^\prime_r {\bf e_\alpha}\right]$ and
 ${\bf E_r}= U\nabla (\varphi_r \cos\alpha)$. The first term ${\bf
E_\alpha}$ here
 is a usual dipole-like field induced around moving vortex. The
 second term ${\bf E_r}$ is the mixed mode contribution which
 exists only in BTRS superconductors.

\begin{figure} [!htb]
 \centerline{\includegraphics[width=1.0\linewidth]{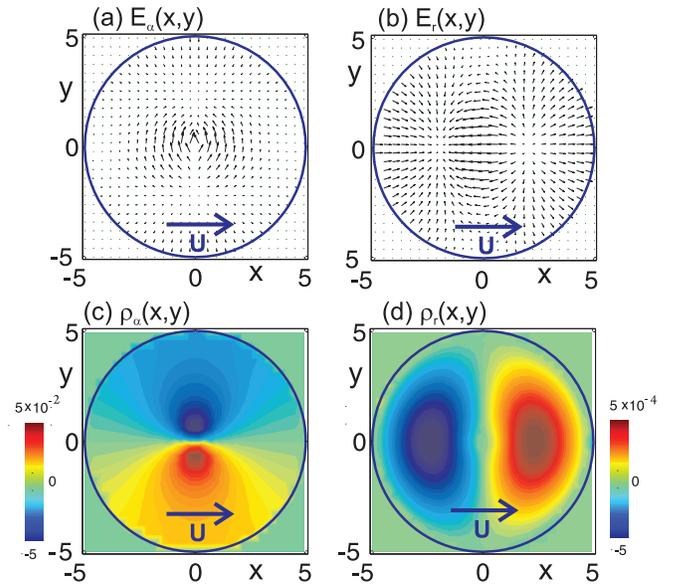}}
 \caption{\label{Fig2}   (a,b) The distributions
 of (a) dipole-like ${\bf E_\alpha}$ and (b) induced by mixed mode ${\bf E_r}$ parts of the total electric field (shown by arrows)
 ${\bf E}={\bf E_\alpha}+{\bf E_r}$  in the unit cell around the vortex moving with velocity ${\bf U}$.
 (c,d) The distribution of electric charge around the moving
 vortex. (c) $\rho_\alpha (x,y) = {\rm div} {\bf E_\alpha}/4\pi$ which coincides with the charge
 in time reversal invariant superconductor. (d) $\rho_r (x,y)={\rm div} {\bf E_r}/4\pi$
 appears in BTRS superconductor.By blue solid circle the boundary of circular cell is shown.}
   \end{figure}

  Distributions of ${\bf E_\alpha}$ and ${\bf E_r}$ components
of the electric field are shown in the Fig.(\ref{Fig2}) a,b. From
Fig.(\ref{Fig2})a one can see that the component ${\bf E_\alpha}$
determines the average electric field in the sample
 $\langle{\bf E}\rangle = (\pi r^2_s)^{-1}\int_{u.c.} {\bf E_\alpha} d^2r =
[{\bf B}\times {\bf U}]$ where ${\bf B}$ is the average magnetic
induction. The other component ${\bf E_r}$ shown in
Fig.(\ref{Fig2}) b does not contribute to the average $\langle{\bf
E_r}\rangle_{u.c.}=0$.

  The relation between vortex velocity ${\bf U}$ and
 transport current ${\bf j_{tr}}$ is in general determined by the
  balance of the forces acting on the moving vortex. There are two of them: Lorentz
force from the transport current  and the force from the
environment ${\bf f_{env}}$ given by the expression (under the
assumption that London penetration length is much larger than the
vortex core size)
 \begin{equation}\label{Fenv}
 {\bf f_{env}} = 2\sum_{k=1}^3\int \Gamma_k \left( \nabla \Delta_k
 \dot{\Delta}_k + \Phi_k {\bf Q_k}
 \Delta_k^2 \right) d^2{\bf r}
\end{equation}
where we introduced gauge invariant scalar potential
$\Phi_k=\varphi_k-\dot{\theta}_k$. { To find a linear response of
the environment we need to keep the
 terms in Eq.(\ref{Fenv}) up to the first order in vortex
 velocity.} In this approximation the force from the environment  provides
viscous drag which has the general form ${\bf f_{env}}=-\eta {\bf
U}$  where $\eta$ is vortex viscosity.  We find that in BTRS
superconductor it can be presented as a
 superposition of three terms of different physical origin
$\eta=\eta_{T}+\eta_{BS}+\eta_{mix}$.
 Here the first two terms appear in ordinary
viscous vortex motion: these are the Tinkham \cite{TinkhamBook}
and
 Bardeen-Stephen \cite{BardeenStephen} contributions.
 The third term $\eta_{mix}$ is completely new and appears due to
 the electric {\it mixed phase-density mode} in the BTRS vortex
 core:
\begin{equation}\label{Eq:visLegg}
 \eta_{mix}= 2\pi \sum_{k=1}^3\Gamma_k \int_0^{\infty}
 \Delta_k^2\left[  r\theta_k^\prime(\theta_k^\prime+\varphi_r)  \right]
 dr
 \end{equation}
 where we put $r_s=\infty$ for well-separated vortices.
 The physical origin of viscosity (\ref{Eq:visLegg}) is the electric field
 excitation due to the mixed mode. The corresponding electric field pattern
 and charge density around moving vortex is shown in Figs.(\ref{Fig2})b,d.

We calculated the total vortex viscosity $\eta$ by solving
Eqs.(\ref{Eq:phi-alpha},\ref{Eq:phi-r}) using the vortex structure
determined by static GL Eqs.(\ref{Eq:System}). We find a striking
behavior of viscosity near the BTRS transition shown in
Fig.\ref{Fig:order1}d as function of interband coupling
$\gamma_{13}$. The mixed mode contribution (shown by black dashed
line) has a pronounced maximum near BTRS phase transition where
the mixed mode becomes massless. The viscosity however
 still remains finite even at the
critical point due to the anharmonism in Eqs.(\ref{Eq:System})
which provides a power-law decay for the phase-density mode
contribution to viscosity as well as to the vortex energy . The
conventional Tinkham and Bardeen- Stephen contributions are
monotonic functions of $\gamma_{13}$. Summing up all contributions
we find that here the behavior of the total viscosity is dominated
by the mixed mode near the BTRS transition. It is shown by solid
red line which features a pronounced peak. This anomalous behavior
is realized for vortex structures belonging to Branch 2 with BTRS
 either in the bulk at $\gamma_{13}>\gamma^*_{13}$ or in
the vortex core at $\gamma_{13}<\gamma^*_{13}$. On the other hand
vortex solution without BTRS corresponding to the Branch 1 has
monotonic viscosity. Comparing Figs.\ref{Fig:order1}b and
\ref{Fig:order1}d one can see that the vortex energy and viscosity
behave rather similar as functions of $\gamma_{13}$.

 In conclusion we reported a new mechanism contributing to vortex viscosity
in BTRS superconductors.
 The results  are generic for BTRS superconductors with mode mixing.
In particular it is not specific to three-band superconductor but
should also apply to  BTRS states with different number of bands
or different interband frustration which exhibit mode mixing (some
of which were discussed in \cite{weston}). It  leads to a
pronounced anomaly at the phase transition where time reversal
symmetry is broken.
 Thus one can potentially
observe this phase transition
 by measuring the anomalous behavior of both
thermodynamic properties (vortex energy $\varepsilon_v$ determines
the lower critical field $H_{c1}=\varepsilon_v/\Phi_0$) and
transport properties such as flux flow resistance which is
determined by vortex viscosity.  It can be utilized to detect
possible $s+is$ state in $Ba_{1-x}K_{x}Fe_2As_2$.

We thank Daniel Weston for discussions. MS was supported by the
Swedish Research Council, Russian Foundation for Basic Research
Grants No 11-02-00891, 13-02-97126 and Russian President
scholarship (SP- 6811.2013.5), EB was supported by the US National
Science Foundation CAREER Award No. DMR-0955902, and
 by Knut and Alice Wallenberg Foundation through the Royal Swedish Academy of Sciences,
Swedish Research Council.

\section{Supplementary material}
\subsection{Power law asymptotic of order parameter fields in zero mass regime}
We consider the system
\begin{widetext}
 \begin{eqnarray}
  \label{Eq:PhaseSupp}
&g\left[\nabla^2_r \theta + 2 (\ln \Delta)^\prime_r
\theta^\prime_r\right]
   + \frac{\gamma\Delta_2}{\Delta}\sin\theta+
   \gamma_{13}\sin(2\theta)=0&\\
  \label{Eq:Delta1Supp}
&\left[g_2\left(\nabla^2_r-\bf{Q_2^2}\right)-\alpha_2-\beta_2\Delta_2^2\right]\Delta_2=
    2\gamma \Delta \cos\theta & \\ \label{Eq:DeltaSupp}
 &\left[g\left(\nabla^2_r-\bf{Q^2}\right)-\alpha- \gamma_{13}  \cos(2\theta) -\beta\Delta^2\right]\Delta=\gamma\Delta_2
 \cos\theta &
\end{eqnarray}
\end{widetext}
where ${\bf{Q}^2}=(A+1/r)^2+ \theta^{\prime 2}_r$ and
${\bf{Q}}^2_2=(A+1/r)^2$. For the vector potential we use the
radial gauge ${\bf A}=A(r)(-\sin\alpha, \cos\alpha)$.

We are interested in particular case when coupling parameters
satisfy the condition $\gamma_{13}=\gamma_{13}^* =
2\gamma\Delta_{0}/\Delta_{20}$. In this case the mass of phase
density mixed mode is zero $\mu_0(\gamma_{13}^*)=0$ and the
asymptotic of coupled phase density fluctuation far from the
vortex core has power law behavior. We search the deviations of
order parameter density and phase from bulk values in the form of
power law expansion
 \begin{equation}\label{Eq:Expansion}
  (\tilde{\Delta},\tilde{\Delta}_2,\tilde{\theta})=
(C_\Delta/r^{p},C_{\Delta 2}/r^{p_2},C_\theta/r^q)
\end{equation}
 Substituting this ansatz into the system
(\ref{Eq:PhaseSupp},\ref{Eq:Delta1Supp},\ref{Eq:DeltaSupp}) we
require that the lowest order terms have the same dependence on
$1/r$. This condition determines the exponents $p=p_2=2$ and $q=1$
in (\ref{Eq:Expansion}). Furthermore we obtain the linear system
to determine coefficients in Eq.(\ref{Eq:Expansion})
\begin{eqnarray}
 & \frac{\gamma\Delta_{20}}{\Delta_{0}} C_\Delta - \gamma C_{\Delta 2} - \gamma_{13} \Delta_0 C_\theta^2= g\Delta_0 \\
 & (\alpha+3\beta\Delta_0^2+\gamma_{13})C_\Delta - \gamma C_{\Delta 2}- \gamma_{13}\Delta_0 C_\theta^2 = 0 \\
 & 2\gamma C_\Delta - (\alpha_2+3\beta_2\Delta_{20}^2) C_{\Delta 2} - \gamma \Delta_0 C_\theta^2 =
 0.
\end{eqnarray}
 For the parameters employed for numerical calculations we obtain
 $C_\Delta = 0.17\Delta_0$, $C_{\Delta 2}=  -0.18\Delta_{20}$  and $C_\theta=
 1.57$.

\subsection{Vortex structure near the critical point $\gamma_{13}=\gamma^c_{13}$}
 The critical point separates regimes in $U(1)$ region with single
 and double solutions for the vortex structure. The solution with
 spatial variation of relative phase continuously emerges at
 $\gamma_{13}>\gamma^c_{13}$ where $\gamma^c_{13}$ is given
 by the eigenvalue of linear equation which can be written in the
 form
  \begin{equation} \label{Eq:PhaseLinSupp}
   \hat L \tilde{\theta}=
   2\gamma^c_{13} \Delta^2 \tilde{\theta}
 \end{equation}
 where $\tilde{\theta}=\pi-\theta$ and
\begin{equation}\label{Eq:OperatorSupp}
\hat L= -g\left[\Delta^2\nabla^2_r  + (\Delta^2)^\prime_r
\partial_r\right] + \gamma\Delta_2\Delta
\end{equation}
is a { hermitian} operator and therefore has orthogonal
eigenfunctions. This form allows to find approximate solution of
nonlinear Eq.(\ref{Eq:PhaseSupp}) for small values of
$\gamma^c_{13}-\gamma_{13}$.

  We search the solution of nonlinear
 Eq.(\ref{Eq:PhaseSupp}) in the form $\tilde{\theta}= A \theta_{lin}+\Theta$
 where $\theta_{lin}=\theta_{lin} (r)$ is the
 normalized eigenfunction of Eq.(\ref{Eq:PhaseLinSupp}) and
 $\Theta=\Theta(r)$ is a small correction.
 It collects the contribution of higher levels of the operator (\ref{Eq:OperatorSupp}) and therefore is orthogonal to
 $\theta_{lin} (r)$ so that
\begin{equation}\label{Eq:OrthSupp}
\int_0^{\infty} r \Delta^2 \theta_{lin}  \Theta dr = 0
\end{equation}

 To determine the amplitude $A$ we rewrite Eq.(\ref{Eq:PhaseSupp}) in the form
\begin{equation}\label{Eq:CorrSupp}
\hat L \Theta  = 2(\gamma^c_{13}-\gamma_{13})\Delta^2 A
\theta_{lin} + A^3N(\theta_{lin})
\end{equation}
 where the last term is nonlinear part
 $$
 N(\theta_{lin})=(8\gamma_{13}\Delta^2-\gamma\Delta\Delta_2)\theta_{lin}^3/6.
 $$
 obtained with the help of Taylor expansion $\sin\theta\approx
 -\tilde{\theta}+\tilde{\theta}^3/6$.
 Taking the inner product of both parts of Eq.(\ref{Eq:CorrSupp})
 with $\tilde{\theta}_{lin} (r)$ and employing the hermiticity of
 operator $\hat L$ and orthogonality (\ref{Eq:OrthSupp}) we get
 the amplitude
\begin{equation}\label{Eq:AmplSupp}
 A=\sqrt{\frac{2(\gamma_{13}-\gamma^c_{13})}{\int_0^{\infty} r \theta_{lin}^2 N
 (\theta_{lin})dr}}
\end{equation}

  Thus we obtain that at $\gamma_{13}>\gamma^c_{13}$ the vortex
 structure can have two  solutions. One is that with constant
 interband phase $\theta (r) =const$ and the second one is with
 the phase variation given by the eigenfunction of operator
 (\ref{Eq:OperatorSupp}) with the amplitude
 $A\sim\sqrt{\gamma_{13}-\gamma^c_{13}}$ given by
 Eq.(\ref{Eq:AmplSupp}).

\subsection{Time-dependent Ginzburg-Landau theory and forces acting on moving vortex in three-component superconductor}

 We describe the non-equilibrium process of vortex motion near the critical temperature with
  time-dependent GL model generalized to a two-gap superconductor 
    \begin{equation}        
    \Gamma_k (\partial_t-i \varphi)\psi_k=
    -\delta F/\delta\psi^*_k,
    \label{Eq:fderSupp}
    \end{equation}
  where $j=1,2,3$,  $\varphi$ is the electric potential,
  $\Gamma_k$ are damping constants.
 The expression for the supercurrent is then
  ${\bf j_s}= 2 g_k {\bf  Q}_k \Delta_k^2$ where
  ${\bf  Q}_k=\nabla\theta_k+{\bf A}$
  and for the normal current ${\bf j_n}=\sigma_n {\bf E}$ where
  ${\bf E}=\nabla\varphi +  {\bf \dot{A}}$. Note that
  normal and superconducting current can convert into each other thus they are not
  separately conserved. For the superfluid current we have an
  expression ${\bf j_s}= -\delta F/\delta {\bf A} $ so that
 ${\rm div} {\bf j_s} = i (\psi^*_k\delta F/\delta\psi^*_k  - \psi_k\delta
 F/\delta\psi_k)$. Hence from Eq.(\ref{fder}) we obtain that
  ${\rm div} {\bf j_s}= -2 \Gamma_k\Delta_k^2\Phi_k$ where $\Phi_k=(\varphi-\dot{\theta}_k)$. Taking into account
  the total current conservation  ${\rm div} ({\bf j_s}+{\bf j_n})=0$
  we can get the Poisson equation for quasistationary electric field with the electric charge
  density given by $\rho = -{\rm div} {\bf j_s}/(4\pi\sigma_n)$ so that
 \begin{equation}\label{Eq:PhiSupp}
 \sigma_n{\rm div}{\bf E} =  2 \Gamma_k\Delta_k^2\Phi_k.
 \end{equation}
  Assuming the Coloumb gauge for the vector potential ${\rm
div}{\bf A}=0$ we obtain the Poisson equation
 \begin{equation}\label{Eq:PoissSupp}
 \sigma_n\triangle\varphi = 2 \Gamma_k\Delta_k^2(\varphi-\dot{\theta}_k)
 \end{equation}
 which we employ to calculate the distribution of the
 scalar potential $\varphi$ generated by the moving vortex.

Steady state vortex motion with constant velocity {\bf U} is
determined by the force balance ${\bf f_{env}+f_{L}}=0$ between
Lorentz force ${\bf f_{L}}$ acting on the vortex from external
transport current and force from the environment ${\bf f_{L}}$.

  The force acting on the moving vortices
  is determined by the variation of free energy due to the
  small vortex displacement\cite{KopninBook,Dorsey} $\delta F =- ({\bf f}\cdot {\bf
  d})$. In general the variation of the free energy is
 $$
 \delta F= \int \left [\frac{\delta F}{\delta \psi_k} \delta \psi_k +c.c. +
 \frac{\delta F}{\delta {\bf A}} \delta {\bf A}
 + \frac{{\bf B}}{4\pi} {\rm rot} \delta {\bf A} \right] dV
 $$
 The last two terms here can be found using the identity
 $$
 \frac{\delta F}{\delta {\bf A}} \delta {\bf A}
 + \frac{{\bf B}}{4\pi} {\rm rot} \delta {\bf A} +
 {\rm div} [ {\bf B\times \delta A}]
 =  {\bf j_n}\delta {\bf A}
 $$
 therefore neglecting the surface term
 $$
 \delta F= \int \left [\frac{\delta F}{\delta \psi_k} \delta \psi_k +c.c.+ {\bf j_n}\delta {\bf A} \right]
 dV.
 $$
 Besides the variation of the free energy we take into account the
 interaction of vortices with transport current ${\bf j_{tr}}$ created by the
 external source. It is given by
 $$
 \delta F_{ext}= - \int {\bf j_{tr}} \delta {\bf A} dV
 $$
 According to the conventional procedure we consider the
  variation of the free energy due to the vortex displacement
  described by
 \begin{eqnarray}\label{}
 \delta \psi_k = ({\bf d}\cdot \nabla) \psi_k\\
  \delta {\bf A} = ({\bf d}\cdot \nabla) {\bf A}.
 \end{eqnarray}

 {\bf Lorentz force}
 Now we consider the action of the homogeneous transport
 current ${\bf j_{tr}}$ on vortex. To calculate the force acting on
 vortex we evaluate the energy change due the infinitesimal
 translations of vortex center. Then the elementary work of the external force has
 the form
 $$
 \delta F_{ext}= - \int {\bf j_{tr}} \delta {\bf A} dV
 $$
  where $\delta {\bf A} = ({\bf d \cdot \nabla }) {\bf A}$.
 Now we use the following identities
 $$
 {\rm div } [{\bf j_{tr}} ({\bf d}\cdot {\bf A})]=
 {\bf j_{tr}} [{\bf d}\times {\bf B}] + {\bf j_{tr}} ({\bf d}\nabla) {\bf A}
 $$
 to obtain
 $$
 \delta F_{ext}=  {\bf d} \int[ {\bf j_{tr}}\times {\bf B}]
 dV =  2\pi{\bf d} \cdot ({\bf j_{tr}}\times  {\bf z_v})
 $$
 where ${\bf z_v}$ is the vorticity direction.
 Therefore the force is
 $$
 {\bf f_{L}}=2\pi [{\bf j_{tr}}\times  {\bf z_v}].
 $$

{\bf Force from the environment}
 To calculate the force from the environment we should consider the energy variations due
 to displacement
  $\delta \psi_k= -{\bf d}\cdot \nabla_{\bf r} \psi_k$ and $\delta {\bf A}= -{\bf d}\cdot \nabla_{\bf r} {\bf A}$.
  Then we can make use of Eq.(\ref{Eq:fderSupp}) which results
  $$
  \delta F= \int \left [ \frac{\delta F}{\delta\psi_k} ({\bf d} \nabla \psi_k)
  +c.c.+ {\bf j_n}({\bf d} \nabla) {\bf A} \right] dV
  $$
  Further we use the fact  ${\rm div} {\bf j_s} = i (\psi^*_k\delta F/\delta\psi^*_k  - \psi_k\delta
 F/\delta\psi_k)$ and transform the above equation as follows
 \begin{eqnarray}\label{Eq:Supp1} \nonumber
 \delta F= \int \left [\frac{\delta F}{\delta \psi_k} ({\bf d}\cdot (\nabla-i {\bf A} )) \psi_k
 +c.c.\right]dV-\\
 \int \left[ ({\bf d}\cdot {\bf A}) {\rm div} {\bf j_s}
 +  {\bf j_n}[({\bf d}\cdot \nabla) {\bf A}]\right] dV
 \end{eqnarray}
  The last two terms can be written using the identity
  \begin{eqnarray}\label{Eq:Supp2} \nonumber
    {\bf j_n}[({\bf d}\cdot \nabla) {\bf A}] + ({\bf d}\cdot {\bf A}) {\rm div} {\bf j_n}
   ={\bf d}\cdot [{\bf j_n}\times {\bf B}]
 \end{eqnarray}
  Therefore neglecting the surface term we get
 \begin{equation}\label{Eq:Supp3} \nonumber
 \delta F={\bf d}\cdot \int \left [\frac{\delta F}{\delta \psi_k} (\nabla-i {\bf A}) \psi_k +c.c.
 +  [{\bf j_n}\times {\bf B}]\right] dV
 \end{equation}
 Now let us make use of the Eqs. (\ref{Eq:fderSupp}) to substitute
 \begin{eqnarray}\label{Eq:Supp4} \nonumber
 \int \left [\Gamma_k (\partial_t+i
 \varphi)\psi^*_k (\nabla-i {\bf A}) \psi_k +c.c.\right] dV \\
 = 2\int \left [\Gamma_k \left(
 \nabla \Delta_k \dot{\Delta}_k + \Phi_k {\bf Q_k}
 \Delta_k^2 \right)\right] dV.
 \end{eqnarray}
 which finally yields
  \begin{eqnarray}\label{Eq:Supp5} \nonumber
 &\delta F = \\ \nonumber
 &{\bf d}\cdot \int \left [- 2\Gamma_k \left(
 \nabla \Delta_k \dot{\Delta}_k + \Phi_k {\bf Q_k}
 \Delta_k^2 \right) +  [{\bf j_n}\times {\bf B}]\right] dV
 \end{eqnarray}
  For the typical type-II superconductors the last term is usually
  neglected. Then we obtain the force acting on the unit length of moving vortex
  line from the environment
  \begin{equation}\label{FenvSupp}
 {\bf f_{env}} = 2\int \Gamma_k \left( \nabla \Delta_k
 \dot{\Delta}_k + \Phi_k {\bf Q_k}
 \Delta_k^2 \right) d^2r
\end{equation}


\begin{thebibliography}{99}
\bibitem{remm2}
{In the phase-only London model this divergence was identified
earlier as the massless Leggett mode in \cite{Lin}}.
\bibitem{remark}
The existence of a diverging length scale at the $Z_2$ phase
transition and
 mixing of phase difference and density
modes  dictates that there should be diverging coherence length at
the TRSB phase transition while magnetic field penetration length
should stay finite. This in turns implies that if the system
should generically have ``type-1.5" regime  \cite{Johan} in the
vicinity of TRSB state with one coherence length larger and other
smaller than the magnetic field penetration length (if the system
is not a type-I superconductor).
 Here we consider the case of vortex lattice (e.g.
large fields). Results should also apply to interiour of
macroscopically large vortex clusters.
\bibitem{remm}
The stability analysis, requires relaxing axially symmetric
ansatz. It is beyond the scope of this paper and will be presented
elsewhere.
\end{thebibliography}
\end{document}